\documentclass[10pt,letterpaper]{article}
\usepackage{ccn}
\usepackage{pslatex}
\usepackage{apacite}
\usepackage{amsmath}
\usepackage[nosetpagesize]{graphicx}

\title{Tracking Naturalistic Linguistic Predictions with Deep Neural Language Models}

 \author{{\large \bf Micha Heilbron}, 
 {\large \bf Benedikt Ehinger},
 {\large \bf Peter Hagoort} 
 \bf {\&} {\large \bf Floris P. de Lange } \\
  { \large \{m.heilbron, b.ehinger, floris.delange, peter.hagoort\}@donders.ru.nl}
 \\ \\
  Donders Centre for Cognitive Neuroimaging, Radboud University 
Nijmegen, The Netherlands}

\begin{document}

\maketitle

\section{Abstract}
{\bf 
Prediction in language has traditionally been studied using simple designs in which neural responses to expected and unexpected words are compared in a categorical fashion. 
However, these designs have been contested as being `prediction encouraging', potentially exaggerating the importance of prediction in language understanding.
A few recent studies have begun to address these worries by using model-based approaches to probe the effects of linguistic predictability in naturalistic stimuli (e.g. continuous narrative).
However, these studies so far only looked at very local forms of prediction, using models that take no more than the prior two words into account when computing a word's predictability.
Here, we extend this approach using a state-of-the-art neural language model that can take  roughly 500 times longer linguistic contexts into account.
Predictability estimates from the neural network offer a much better fit to EEG data from subjects listening to naturalistic narrative than simpler models, and reveal  strong surprise responses akin to the P200 and N400.
These results show that predictability effects in language are not a side-effect of simple designs, and demonstrate the practical use of recent advances in AI for the cognitive neuroscience of language. 
}
\begin{quote}
\small
\textbf{Keywords:} 
prediction; language; Transformer; GPT-2 
\end{quote}

\section{Introduction}

In a typical conversation, listeners perceive (or produce) about 3 words per second. 
It is often assumed that prediction offers a powerful way to achieve such rapid processing of often-ambiguous linguistic stimuli.
Indeed, the widespread use of language models -- models computing the probability of upcoming words given the previous words -- in speech recognition systems demonstrates the in-principle effectiveness of prediction in language processing \cite{jurafsky_speech_2014}.  

Linguistic predictability has been shown to modulate fixation durations and neural response strengths, suggesting that the brain may also use a predictive strategy. 
This dovetails with more general ideas about predictive processing \cite{friston_theory_2005,de_lange_how_2018,heilbron_great_2017} and has lead to predictive interpretations of classical phenomena like the N400 \cite{rabovsky_modelling_2018,kuperberg_what_2016}.
However, most neural studies on prediction in language used hand-crafted stimulus sets containing many highly expected and unexpected sentence endings -- often with tightly controlled (predictable) stimulus timing to allow for ERP averaging. 
These designs have been criticised as `prediction encouraging' \cite{huettig_is_2016}, potentially distorting the importance of prediction in language.

A few recent studies used techniques from computational linguistics combined with regression-based deconvolution to estimate predictability effects on neural responses to naturalistic, continuous speech. 
However, these pioneering studies probed very local forms of prediction by quantifying word predictability based on only the first few phonemes \cite{brodbeck_rapid_2018} or the prior two words \cite{willems_prediction_2016,armeni_frequency-specific_2019}. 
Recently, the field of artificial intelligence has seen major improvements in neural language models that  predict the probability of an upcoming word based on a variable-length and (potentially) arbitrarily-long prior context. 
In particular, self-attentional architectures \cite{vaswani_attention_2017} like GPT-2 can keep track of contexts of up to a thousand words long, significantly improving the state of the art in long-distance dependency language modelling tasks like LAMBADA and enabling the model to generate coherent texts of hundreds of words \cite{radford_language_2019}.
Critically, these pre-trained models can achieve state-of-the art results on a wide variety of tasks and corpora without any fine-tuning. 
This stands in sharp contrast to earlier (ngram or recurrent) language models which were trained on specific tasks or linguistic registers (e.g. fiction vs news).
As such, deep self-attentional language models do not just coherently keep track of long-distance dependencies, but also exhibit an unparalleled degree of \emph{flexibility}, making them arguably the closest approximation of a `universal model of English'  so far. 

Here we use a state-of-the art pre-trained neural language model (GPT-2 M) to generate word-by-word predictability estimates of a famous work of fiction, and then regress those predictability estimates against publicly-available EEG data of participants listening to a recording of that same work.

\begin{figure*}
\begin{center}
\vspace{-0.2in}
\includegraphics[width=0.8\textwidth]{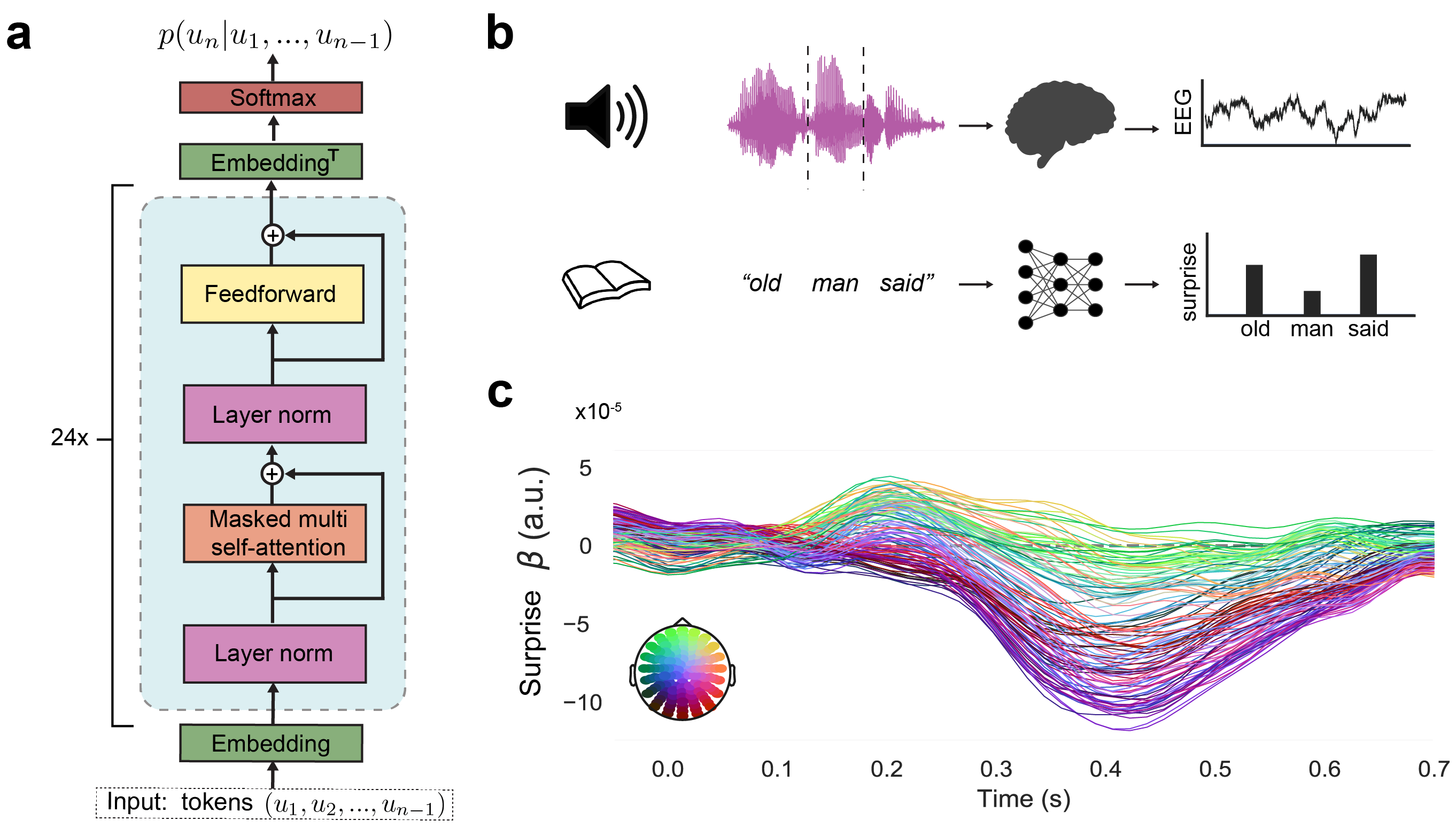}
\end{center}
\caption{{\bf a) } GPT-2 architecture. For more info on individual operations, see Vaswani et al. \citeyear{vaswani_attention_2017}.
(Note that this panel is a re-rendered version of the original GPT schematic, with subcomponents re-arranged to match the architecture of GPT-2.) {\bf b) } Analysis pipeline overview. {\bf c)} Obtained series of $\beta$ coefficients (TRF) of lexical surprise (from GPT-2), averaged over participants.   } 
\label{fig:overview}
\end{figure*}

\section{Methods}

\subsection{Stimuli, data acquisition and preprocessing}
We used publicly available EEG data of 19 native English speakers listening to Hemingway's \emph{The Old Man and the Sea}.
Participants listened to 20 runs of 180s long, amounting to the first hour of the book (11,289 words, $\sim$3 words/s).  
Participants were instructed to maintain fixation and minimise all motor activities but were otherwise not engaged in any task. 

The dataset contains raw 128-channel EEG data downsampled to 128 Hz, plus on/offset  times of every content word.
The raw data was visually inspected to identify bad channels,  decomposed using ICA to remove blinks, after which the rejected channels were interpolated using MNE-python. 
For all analyses, we focussed on the slow dynamics by filtering the z-scored, cleaned data between 0.5 and 8 Hz using a bidirectional FIR. 
This was done to keep the analysis close to earlier papers using the same data to study how EEG tracks acoustic and linguistic content of speech; but note that changing the filter parameters does not qualitatively change the results. 

For more information on the dataset and prior analyses,  see \cite{broderick_electrophysiological_2018}.

\subsection{Computational models}
Word-by-word unpredictability was quantified via lexical surprise -- or $-\log\big(p(word|context)\big)$ -- estimated by GPT-2 and by a trigram language model. 
We will describe each in turn. 


\subsubsection{GPT-2}
GPT-2 is a decoder-only variant of the \emph{Transformer} \cite{vaswani_attention_2017}.
In the network, input tokens $U = (u_{i-k},...,u_{i-1})$ are passed through a token embedding matrix $W_e$ after which a position embedding $W_p$ is added to obtain the first hidden layer: $h_0 = UW_e +W_p$. 
Activities are then passed through a stack of transformer blocks, consisting of a multi-headed self attention layer, a position-wise feedforward layer, and layer normalisation (Fig \ref{fig:overview}a). 
This is repeated $n$ times for each block $b$, after which (log)probabilities are obtained from a (log)softmax over the transposed token embedding of $h_n$:   
\begin{align}
h_{b} &=\operatorname{transformer\_block}\left(h_{b-1}\right) \forall i \in[1, n] \\ 
P(u_i |U) &=\operatorname{softmax}\left(h_{n} W_{e}^{\top}\right)  \label{eq:trans}
\end{align}

We used the largest public version of GPT-2 (345M parameter, released May 9)\footnote{For more details on GPT-2, see https://openai.com/blog/better-language-models/ or Radford et al \citeyear{radford_language_2019}}
 which has a number of layers (blocks) of $n=24$ and a context length of $k=1024$. 
Note that $k$ refers to the number of Byte-Pair Encoded \emph{tokens}.
A token can be either a word or (for less frequent words) a word-part, or punctuation. 
How many words actually fit into a context window of length $k$ therefore depends on the text.
We ran predictions on a run-by-run basis -- each containing about 600 words, implying that in each run the entire preceding context was taken into account to compute a token's probability. 
For words spanning multiple tokens, word probabilities were simply the joint probability of the tokens obtained via the chain rule. 
The model was implemented in PyTorch with the Huggingface BERT module\footnote{see https://github.com/huggingface/pytorch-pretrained-BERT}.  

 \subsubsection{Trigram} 
 
 As a comparison, we implemented an n-gram language model. 
 N-grams also compute $p(w_{i}| w_{i-k},...,w_{i-1})$ but are  simpler as they are based on counts. 
 Here we used a trigram ($k=2$) -- which was perhaps the most widely used language model before the recent rise of neural alternatives.\footnote{While $k=2$ might seem needlessly restrictive,  training ngrams beyond $k=2$ becomes exponentially difficult due to sparsity issues.}
To deal with sparsity we used modified Knesner-Ney,  the best-performing smoothing technique \cite{jurafsky_speech_2014}. 
 The trigram was implemented in NLTK and trained on its Gutenberg corpus, chosen to closely approximate the test set. 
 
\begin{figure*}
\begin{center}
\vspace{-0.2in}
\includegraphics[width=0.8\textwidth]{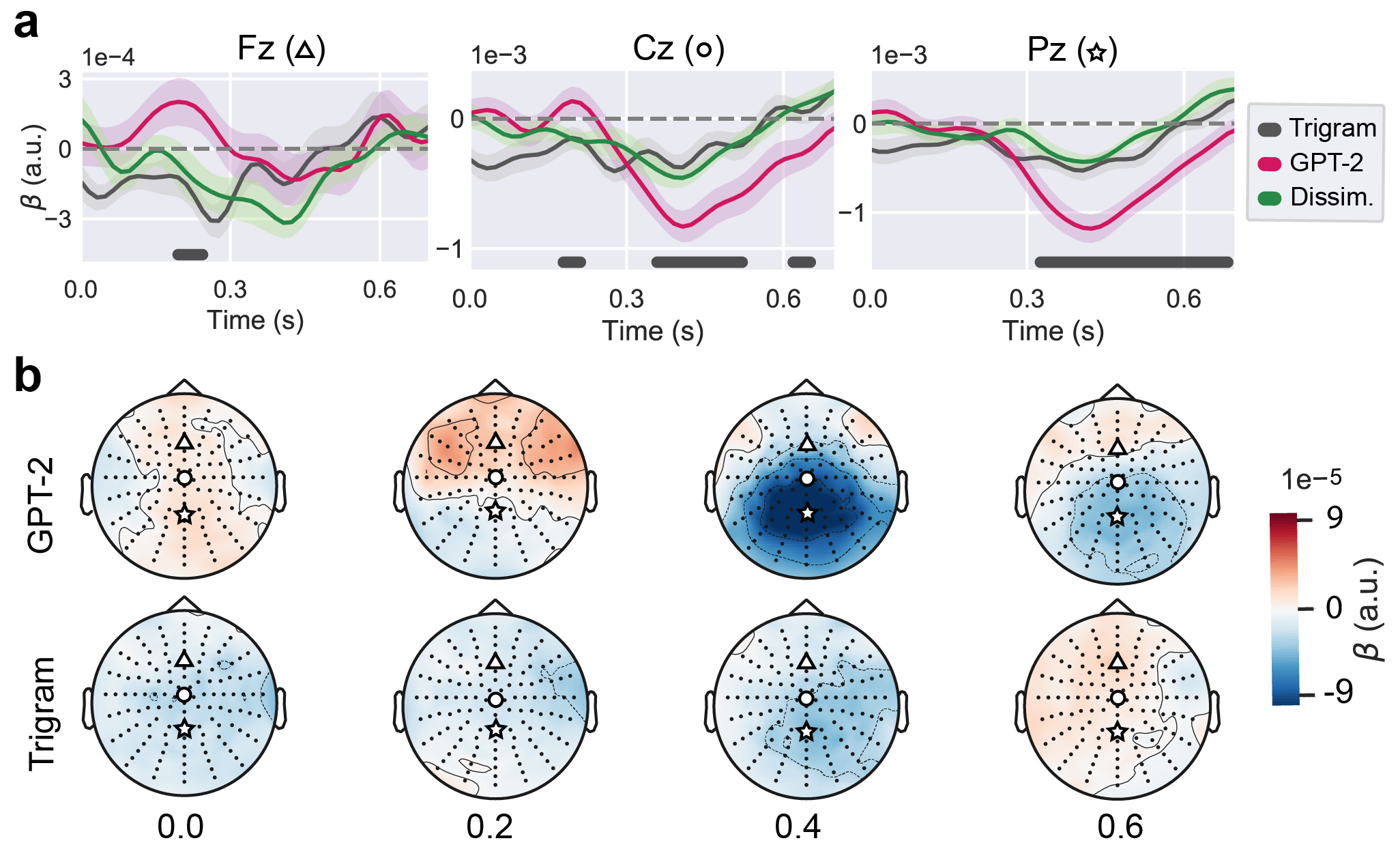}
\end{center}
\caption{{\bf a) } Grand averaged TRFs for trigram surprise,GTP-2 surprise and semantic dissimilarity for three channels of interest.
At each time point, the GPT-2 TRF was compared to both the trigram and semantic dissimilarity TRF with a 2-tailed paired t-test; black bars indicating that both tests were significant at $p<0.01$, FDR-corrected. Error bars indicate the between-subject SEM. 
 {\bf b)} Topographic maps of grand averaged TRFs for surprise, computed by GPT-2 (top) and the trigram language model (bottom). 
}
\label{fig:main_res}
\end{figure*}

 \subsubsection{Non-predictive controls}
 We included two non-predictive and potentially confounding variables: first, frequency which we quantified as unigram surprise ($-\log{p(w)}$) which was based on a word's lemma count in the CommonCrawl corpus, obtained via spaCy. 
 Second, following Broderick et al. \citeyear{broderick_electrophysiological_2018}, we computed the semantic dissimilarity for each content word: $\operatorname{dissim}(w_i)=1-\text{corr}\big(\text{GloVe}(w_i),\frac{1}{n}\sum_{i=i}^{n}{\text{GloVe}(c_i)}\big)$, where $(c_1,...,c_n)$ are the content words preceding a word in the same or -- if $w_i$ is the first content word of the sentence -- the previous sentence, and  $\text{GloVe}(w)$ is the embedding.
As shown by Broderick et al. \citeyear{broderick_electrophysiological_2018} this variable covaries with an N400-like component.
However, it only captures how semantically dissimilar a word is from the preceding words (represented as an `averaged bag of words'), and not how unexpected a word is in its context, making it an interesting comparison, especially for predictive interpretations of the N400.
\subsection{Time resolved regression}
 Variables were regressed against EEG data using time-resolved regression. 
 Briefly, this involves temporally expanding a design matrix such that each predictor column $C$ becomes a series of columns over a range of lags $C_{t_{min}}^{t_{max}}=(C_{t_{min}},...,C_{t_{max}})$. 
For each predictor one thus estimates a series of weights $\beta_{t_{min}}^{t_{max}}$ (Fig \ref{fig:overview}c)  which, under some assumptions, corresponds to the isolated ERP that would have been obtained in an ERP paradigm.
In all analyses, word onset was used as time-expanded intercept and other variables as covariates. 
All regressors were standardised and coefficients were estimated with Ridge regression. Regularisation was set at $\alpha=1000$ since this lead to the highest $R^2$ in a leave-one-run-out CV procedure (Fig. \ref{fig:r2s})
Analyses were performed using custom code adapted from MNE's $\operatorname{linear\_regression}$ module.

\section{Results}

We first inspected our main regressor of interest: the surprise values computed by GPT-2, estimated with a regression model that included frequency (unigram surprise) and semantic dissimilarity as nuisance covariates.
As can be seen in Figure \ref{fig:overview}C, the obtained TRF revealed a clear frontal positive response around 200 ms and a central/posterior negative peak at 400 ms after word onset.
These peaks indicate that words that were more surprising to the network tended to evoke stronger positive responses at frontal channels at 200 ms and stronger negative potentials at central/posterior channels 400 ms after word onset. 
Note that while Figure \ref{fig:overview}C only shows the TRF obtained using one regularisation parameter, we found the same qualitative pattern for any alpha we tested. 

We then compared this to an alternative regression model, in which the surprise regressor was based on the trigram model, but that was otherwise identical. 
Although the TRFs exhibited the same negativity at 400 ms, it was a lot weaker overall, as can be seen from Figure \ref{fig:main_res}B. 
One anomalous feature is that the TRF is not at 0 at word onset. 
We suspect this is because 1) we only had onset times for content words, and not for function words typically preceding content words; and 2) for neighbouring words the log-probabilities from the trigram model were correlated ($\rho=0.24$) but those from GPT-2 were not ($\rho=-0.002$), explaining why only the trigram TRF displays a baseline effect.
Further analyses incorporating onset times for all words should correct this issue.  

The negative surprise response at 400ms revealed by both the trigram and GPT is similar to the effect of semantic dissimilarity reported by Broderick et al. \citeyear{broderick_electrophysiological_2018} using the same dataset.  
We therefore also looked at the TRF of semantic dissimilarity, for  simplicity focussing  on the three main channels of interest analysed by Broderick et al. \citeyear{broderick_electrophysiological_2018}. 
At each time-point we compared the GPT-2 TRF to both the trigram and semantic dissimilarity TRF with a 2-tailed paired t-test to find time-points where both tests where significant at $\alpha=0.01$ (FDR-corrected). 
As visible in Figure \ref{fig:main_res}b, we observed timepoints in all three channels where the GPT-2 TRF was significantly more positive or negative than both other TRFs, confirming that the surprise values from the neural network covary more strongly with EEG responses than the other models.

\begin{figure}
\begin{center}
\vspace{-0.35in}
\includegraphics[width=0.25\textwidth]{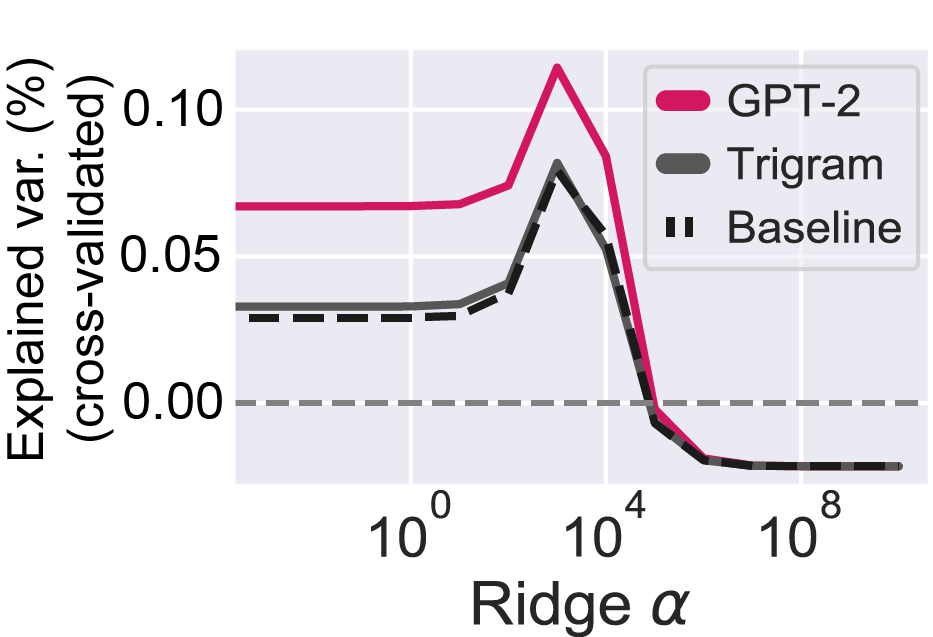}
\end{center}
\vspace{-.15in}
\caption{Predictive performance of three regression models. We compared a baseline regression model with only unigram surprise and semantic dissimilarity as covariates (dotted line) to two other  models that also included surprise values, either obtained from the trigram model (grey) or from GPT-2 (red). } 
\vspace{-.1in}
\label{fig:r2s}
\end{figure}

Finally, to make sure that the difference in coefficients were not related to overfitting or some other estimation problem, we compared the predictive performance of the GPT-2 regression model to the alterntives using a leave-one-run-out cross-validation procedure. 
As can be seen in Figure \ref{fig:r2s}, this revealed that cross-validated $R^2$ of the trigram regression model was not significantly higher than that of a baseline model that included only the two nuisance covariates (paired t-test, $t_{19}=-0.25, p=0.8$); by contrast, $R^2$ of the GPT-2 regression model was significantly higher than both the trigram regression model (paired t-test, $t_{19}=5.38,p=4.1\times10^{-4}$) and the baseline model (paired t-test, $t_{19}=3.10§,p=6.2\times10^{-3}$).

\section{Discussion and conclusion}

We have shown that word-by-word (un)predictability estimates obtained with a state-of-the-art self-attentional neural language model systematically covary with evoked brain responses to a naturalistic, continuous narrative, measured with EEG. 
When this relationship was plotted over time, we observed a frontal positive response at 200 ms, and a central negative response at 400 ms, akin to the N400. 
Unpredictability estimates from the neural network were a much better predictor of EEG responses than those obtained from a trigram that was specifically trained on works of fiction, and than a non-predictive model of semantic incongruence, that simply computed the dissimilarity between a word and its context. 

These results bear strong similarities to earlier work demonstrating a relationship between the N400 and semantic expectancy. 
However, we observed the responses in participants passively listening to naturalistic stimuli, without many highly expected or unexpected sentence endings typically used in the stimulus sets of traditional ERP studies. 
This suggests that linguistic predictability effects are not just a by-product of simple (prediction encouraging) designs, underscoring the importance of prediction in language processing.

Future analyses will aim at modelling all words, looking at different frequency bands, disentangling different forms of linguistic prediction (e.g. syntactic vs semantic), and trying to replicate these results in different, independent datasets.

\section{Acknowledgments}

We want to thank Michael Broderick and the Lalor lab for sharing the data, and all authors of open source software we used. 
This work was supported by NWO (Vidi grant to FdL, Research Talent grant to MH), the James S. McDonell Foundation (JSMF scholar award to FdL), and the EU Horizon 2020 Program (ERC starting grant 678286 to FdL). 


\bibliographystyle{apacite}
\setlength{\bibleftmargin}{.125in}
\setlength{\bibindent}{-\bibleftmargin}

\bibliography{CCN_mh.bib}

\end{document}